\documentclass[usenatbib,usegraphicx]{mn2e}
\def\msun{{\rm M}_{\odot}}
\def\rsun{{\rm R}_{\odot}}

\def\kms{\rm km\ s^{-1}}
\def\angstrom{\rm \AA}

\begin{document}

\title[A $0.26$, $0.15\ \msun$ pre-main-sequence eclipsing binary in
    the Orion Nebula Cluster]{The Monitor project: JW 380 -- a $0.26$,
    $0.15\ \msun$ pre main sequence eclipsing binary in the Orion
    Nebula Cluster}
\author[J.~Irwin et al.]{Jonathan~Irwin$^{1}$, Suzanne~Aigrain$^{1,2}$,
    Simon~Hodgkin$^{1}$, Keivan~G.~Stassun$^{3}$,
\newauthor
Leslie~Hebb$^{4}$, Mike~Irwin$^{1}$, Estelle~Moraux$^{5}$,
    Jerome~Bouvier$^{5}$, Aude~Alapini$^{2}$,
\newauthor
Richard~Alexander$^{6}$, D.M.~Bramich$^{7}$, Jon~Holtzman$^{8}$,
    Eduardo~L.~Mart\'{i}n$^{9}$,
\newauthor
Mark~J.~McCaughrean$^{2}$, Fr\'{e}d\'{e}ric Pont$^{10}$,
    P.E.~Verrier$^{1}$, Mar\'{i}a~Rosa~Zapatero~Osorio$^{9}$
 \\
$^{1}$Institute of Astronomy, University of Cambridge, Madingley Road,
  Cambridge, CB3 0HA, United Kingdom \\
$^{2}$Astrophysics Group, School of Physics, University of Exeter, Stocker Road,
  Exeter, EX4 4QL, United Kingdom \\
$^{3}$Department of Physics and Astronomy, Vanderbilt University, VU
    Station B 1807, Nashville, TN 37235, USA \\
$^{4}$School of Physics and Astronomy, University of St Andrews,
  North Haugh, St Andrews, KY16 9SS, Scotland \\
$^{5}$Laboratoire d'Astrophysique, Observatoire de Grenoble, BP 53,
  F-38041 Grenoble C\'{e}dex 9, France \\
$^{6}$JILA, University of Colorado, Boulder, CO 80309-0440, USA \\
$^{7}$Isaac Newton Group of Telescopes, Apartado de Correos 321,
  E-38700 Santa Cruz de la Palma, Canary Islands, Spain \\
$^{8}$Astronomy Department, New Mexico State University, Las Cruces,
    NM 88003, USA \\
$^{9}$Instituto de Astrof\'{i}sica de Canarias, C/ V\'{i}a L\'{a}ctea,
    s/n E38205, La Laguna, Tenerife, Spain \\
$^{10}$Observatoire Astronomique de l'Universit\'{e} de Gen\`{e}ve, 51
    chemin des Maillettes, CH-1290 Sauverny, Switzerland}

\date{}

\maketitle

\begin{abstract}
We report the discovery of a low-mass ($0.26 \pm 0.02$, $0.15 \pm
0.01\ \msun$) pre-main-sequence eclipsing binary with a $5.3\ {\rm
  day}$ orbital period.  JW 380 was detected as part of a high-cadence
time-resolved photometric survey (the Monitor project) using the $2.5\
{\rm m}$ Isaac Newton Telescope and Wide Field Camera for a survey of
a single field in the Orion Nebula Cluster (ONC) region in $V$ and $i$
bands.  The star is assigned a $99$ per cent membership probability
from proper motion measurements, and radial velocity observations
indicate a systemic velocity within $1 \sigma$ of that of the
ONC. Modelling of the combined light and radial velocity curves of the
system gave stellar radii of $1.19^{+0.04}_{-0.18}\ \rsun$ and
$0.90^{+0.17}_{-0.03}\ \rsun$ for the primary and secondary, with a
significant third light contribution which is also visible as a third
peak in the cross-correlation functions used to derive radial
velocities.  The masses and radii appear to be consistent with stellar
models for $2-3\ {\rm Myr}$ age from several authors, within the present 
observational errors.  These observations probe an important region of
mass-radius parameter space, where there are currently only a handful
of known pre-main-sequence eclipsing binary systems with precise
measurements available in the literature.
\end{abstract}
\begin{keywords}
stars: individual: JW 380 -- open clusters and associations:
individual: Orion Nebula Cluster -- binaries: eclipsing -- stars:
pre-main-sequence -- surveys.
\end{keywords}

\section{Introduction}
\label{intro_section}

Detached eclipsing binaries (EBs) provide one of the most precise
($\la 2$ per cent) and accurate (largely model and distance
independent) methods for measurement of fundamental stellar
properties (particularly, masses and radii).  These can be used to
place stringent constraints on stellar evolution models.

On the pre-main-sequence (PMS), such constraints are presently
extremely scarce below $1\ \msun$, and to our knowledge there are only
$5$ known PMS EB systems in this mass range.  These are a $1.0$, $0.7\
\msun$ EB \citep{st2004} and a $1.27$, $0.93\ \msun$ EB
(\citealt{cov2004}; \citealt{cov2001}), both in Orion (the former is
thought to belong to the Ori 1c association), with corresponding age
$\sim 5 - 10\ {\rm Myr}$, an M-dwarf EB in NGC 1647 ($\sim 150\ {\rm
  Myr}$; \citealt{hebb2006}), and two EBs in the ONC: a brown
dwarf-brown dwarf system (\citealt*{st2006}; \citealt*{st2007}) and a
$0.4$, $0.4\ \msun$ M-dwarf EB (Cargile, Stassun \& Mathieu,
submitted).  Comparison of the NGC 1647 and ONC brown dwarf systems to
a variety of stellar models (specifically: \citealt{bcah98};
\citealt*{siess97}; \citealt{yi2001}; \citealt{gir2000} for the
former, and \citealt{bcah98}; \citealt{burrows97}; \citealt{dm97} for
the latter) by the respective authors has indicated that none fit both
components of the binaries simultaneously.  In the case of the brown
dwarf EB, the models do seem to be reasonably consistent with both
objects, but the secondary appears to be hotter than the primary, a
very surprising result that was not predicted by any of the models.

The study of low-mass stars poses a challenge for stellar models.
Stars near to the hydrogen burning limit are sufficiently cool that
the interior is in a partially-degenerate state on the main sequence
\citep{cb97}, and magnetic fields may play an important role
\citep{mm2001}.  The approximations underlying the usual `grey'
atmosphere models break down, so non-grey model atmospheres must be
determined, taking account of effects such as the recombination of
molecules (e.g. H$_2$ and TiO) due to the low temperatures
\citep{bcah2002}.  One of the only ways to test these models is by
comparison with precise observations of low-mass stars.

It is therefore essential that a larger sample of pre-main sequence
systems be found and characterised, to provide better constraints on
the models. This is the primary aim of the Monitor project
(\citealt{a2007}; \citealt{hodg06}), a photometric monitoring survey
we have undertaken of all suitable nearby, pre-main sequence open
clusters and star forming regions, to search for detached eclipsing
binary and transiting planet systems.  This publication presents the
first of these, a detached EB in the Orion Nebula Cluster, which has
age $1 \pm 1\ {\rm Myr}$ and distance $\sim 470\ {\rm pc}$
\citep{h97}.

Sections \ref{phot_section} and \ref{spec_section} review the
photometric and spectroscopic observations and analysis,
respectively.  Section \ref{lcanal_section} presents the light curve
analysis and system parameters, and in Section \ref{modelcomp_section}
these are compared with the predictions of stellar models.  Finally,
we summarise our conclusions in \S \ref{concl_section}.

\section{Photometry}
\label{phot_section}

\subsection{Survey photometry}
\label{origphot_section}

Eclipses in JW 380 were initially detected in our photometric
monitoring data of the ONC obtained using the $2.5\ {\rm m}$ Isaac
Newton Telescope (INT), with the Wide Field Camera (WFC), during two
$10$-night observing runs, one in late-November 2004, and one in
January 2005. This was supplemented by two further $10$-night runs in
December 2005 and 2006.

A single $\sim 34' \times 34'$ field centred on the Trapezium region
($\theta^1_C$ Ori) was observed for the entire time it was at airmass
$< 2.0$, $\sim 6\ {\rm hours}$ per night, using alternating $60\
{\rm s}$ $V$-band and $30\ {\rm s}$ $i$-band exposures.  Due to the
fast ($\sim 40-45\ {\rm s}$) readout of the WFC this gave a cadence of
$\sim 3.5\ {\rm minutes}$.  A total of $1400$ exposures were obtained
in each passband, for a total time on target of $\sim 80\ {\rm
  hours}$.  Our observations are sufficient to give an RMS per 
data point of $1 \%$ or better down to $i \sim 17$ and $V \sim 18$.

Light curves were extracted for a total of $2500$ objects using our
differential photometry software \citep{i2007}.  The $V$-band light
curves in particular are somewhat affected by the presence of
nebulosity in the ONC region, and show significantly more scatter than
the $i$-band data.  We therefore used the latter for detection of
eclipsing binary systems, and the $V$-band observations for
confirmation.

\begin{figure*}
\centering
\includegraphics[angle=270,width=6in]{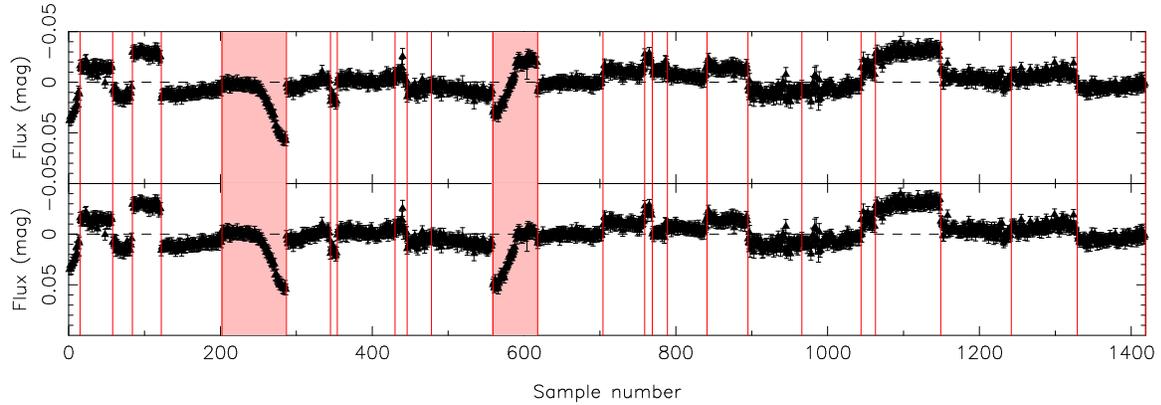}

\caption{INT/WFC $i$-band light curve, plotted as a function of sample
  number (numbering from one for the first observation).  Vertical
  lines denote the boundaries between different nights.  The
  horizontal dashed line indicates the median light curve level.  The
  upper panel shows the data, and the lower panel the data with the
  median light curve level subtracted from nights containing eclipses
  (the nights which were modified are shaded in the figure),
  to reduce the effect of the out-of-eclipse variations on the
  measured eclipse depths.}
\label{290_lc}
\end{figure*}

Due to the extensive intrinsic stellar variability seen in ONC stars
(e.g. \citealt{smmv99}, \citealt{h2002}), the search for eclipsing
systems was by necessity performed manually.  The first candidate
detected in this way is the subject of the present paper.  The
$i$-band light curve (see the upper panel of Fig. \ref{290_lc}) shows
$\sim 0.05\ {\rm mag}$ eclipse events, with $\sim 0.03\ {\rm mag}$
peak-to-peak out of eclipse variations.

The object appears in the catalogue of \citet{jw88} as star $380$,
with membership probability $0.99$ derived from proper motion
measurements by these authors.  It was also detected by the Chandra
Orion Ultradeep Project (COUP; \citealt{getman2005}, star 468), and in
the Two Micron All-Sky Survey (2MASS), as source 2MASS
J05351214-0531388.

The photometric properties of the object are summarised in Table
\ref{290_phot}.  $V$ and $I$ magnitudes were derived from the Monitor
photometry using colour equations from the INT wide field survey web
pages\footnote{\tt
  http://www.ast.cam.ac.uk/\~{}wfcsur/technical/photom/colours/}, and
the near-IR $J$, $H$ and $K$ magnitudes from the 2MASS photometry
using colour equations in the 2MASS explanatory
supplement\footnote{\tt 
  http://www.ipac.caltech.edu/2mass/releases/allsky/doc/sec6\_4b.html}.  Spitzer/IRAC measurements are from Table 1 of \citet{rebull06}.

\begin{table}
\centering

\begin{tabular}{lrl}
\hline
\hline
Passband      &Magnitude         &Source \\
\hline
$V_J$         &$16.92$           &INT/WFC \\
$I_C$         &$13.82$           &INT/WFC \\
$J_{\rm CIT}$ &$12.13 \pm 0.02$  &2MASS\\
$H_{\rm CIT}$ &$11.42 \pm 0.03$  &2MASS\\
$K_{\rm CIT}$ &$11.14 \pm 0.02$  &2MASS\\
$[3.6]$       &$10.86 \pm 0.005$ &Spitzer/IRAC\\
$[4.5]$       &$10.80 \pm 0.005$ &Spitzer/IRAC\\
$[5.8]$       &$10.72 \pm 0.033$ &Spitzer/IRAC\\
$[8.0]$       &$10.04 \pm 0.086$ &Spitzer/IRAC\\
\hline
\hline
\end{tabular}

\begin{tabular}{lrl}
\hline
\hline
Passbands                   &Colour            &Source \\
\hline
$V_J - I_C$                 &$3.10$            &INT/WFC \\
$J_{\rm CIT} - H_{\rm CIT}$ &$0.71 \pm 0.04$   &2MASS\\
$H_{\rm CIT} - K_{\rm CIT}$ &$0.28 \pm 0.04$   &2MASS\\
$[3.6]-[4.5]$               &$0.06 \pm 0.007$  &Spitzer/IRAC\\
$[4.5]-[8.0]$               &$0.76 \pm 0.086$  &Spitzer/IRAC\\
\hline
\hline
\end{tabular}

\caption{Photometric properties of the eclipsing binary system.
  Errors are not quoted for the INT/WFC photometry because these are
  dominated by systematic errors from the sky subtraction due to the
  extensive nebulosity in the optical.  We expect that these are $<
  0.05\ {\rm mag}$.}
\label{290_phot}
\end{table}

The measured composite $I$-band  magnitude and the models of
\citet{bcah98} imply a total system mass of $\sim 0.5\ \msun$ if the
system was a single star, or assuming that the primary contributes
half of the system light, $\sim 0.3\ \msun$ (the latter is a more
reasonable assumption for a near equal mass binary), and the optical
colour $V - I = 3.1$ implies a spectral type of $\sim$ M5 using the
intrinsic colours of \citet{leg92}.

Examination of the COUP light curve\footnote{Light curves are
  available in the source atlas on the COUP
web site: {\tt http://www.astro.psu.edu/coup/}} shows two flare
events, but no obvious evidence for eclipses at the present time.
This is under investigation, in collaboration with members of the COUP
team.

Spitzer/IRAC measurements for JW 380 are available in Table 1 of
\citet{rebull06}, and are reproduced here in Table \ref{290_phot}.
Comparing the observed $[3.6]-[4.5]$ and $[4.5]-[8.0]$ colours to
those derived in NGC 2362 by \citet{dh07} indicates that these
measurements may provide evidence for a weak disc excess, but this
conclusion is somewhat ambiguous due to the obvious composite nature
of the system, which will affect the observed colours.

From the full INT/WFC data-set, a preliminary period of $2.65\ {\rm
  days}$ was determined using a standard box-fitting least squares
transit search program \citep{ai2004}.

\subsection{Follow-up photometry}
\label{followup_sect}

Initial follow-up observations were obtained during 2006 February by
J. Holtzman using the New Mexico State University $1.0\ {\rm m}$
robotic telescope, in $I$-band, at the predicted times of eclipse from
the INT data.

JW 380 was also monitored using the $0.9$, $1.0$ and $1.3\ {\rm m}$
telescopes\footnote{Now operated by the SMARTS consortium.} at the
Cerro Tololo Inter-American Observatory (CTIO).  Observations were
obtained from 2005 December to 2007 January, in $V$ and $I$ bands,
although the majority of the data are in $I$-band.  Differential light
curves were determined from PSF photometry using an algorithm for
inhomogeneous ensemble photometry \citep{h92} as implemented in
\citet{smmv99}, \citet{st2002} for observations of high-nebulosity
regions such as the ONC.

All the available data were combined to produce two composite light
curves, one in $I$-band and the other in $V$-band, applying a
normalisation to account for zero point offsets between the various
photometric systems in use, based on the median out-of-eclipse level.
Note that the $i$ and $I$ passbands are not strictly identical: the
$i$ (SDSS) passband is slightly bluer than the conventional $I$
(Cousins) passband, by $\sim 10$ per cent in the $V-I$ colour.
However, with the present data it is difficult to correct for this
effect, and we expect that the errors introduced by not doing so will
be negligible given the photometric errors.

An updated period of $2.6496\ {\rm days}$ was determined by applying a
new double trapezoid fitting program (Aigrain et al., in prep).

\subsection{Out of eclipse variations}

The out of eclipse variations were found to not phase-fold at the same
period as the eclipsing binary, and moreover with the full combined
data-set it was not possible to find a consistent period for them,
presumably due to changes in the spot coverage of the stellar surfaces
causing phase and amplitude changes in the out of eclipse modulations,
over shorter time-scales than the observing window.

The INT/WFC data considered alone are concentrated into $4 \times$
$\sim 10\ {\rm night}$ observing runs, so we attempted fitting of the
out-of-eclipse parts of this light curve.  If these are due to spots
on the surface of one of the component stars, this allows the rotation
period of the star to be determined, and if a sufficiently good model
for the spot behaviour can be found, the modulations can be removed,
to improve the accuracy of the eclipsing binary model fit.

\begin{figure*}
\centering
\includegraphics[angle=270,width=6in]{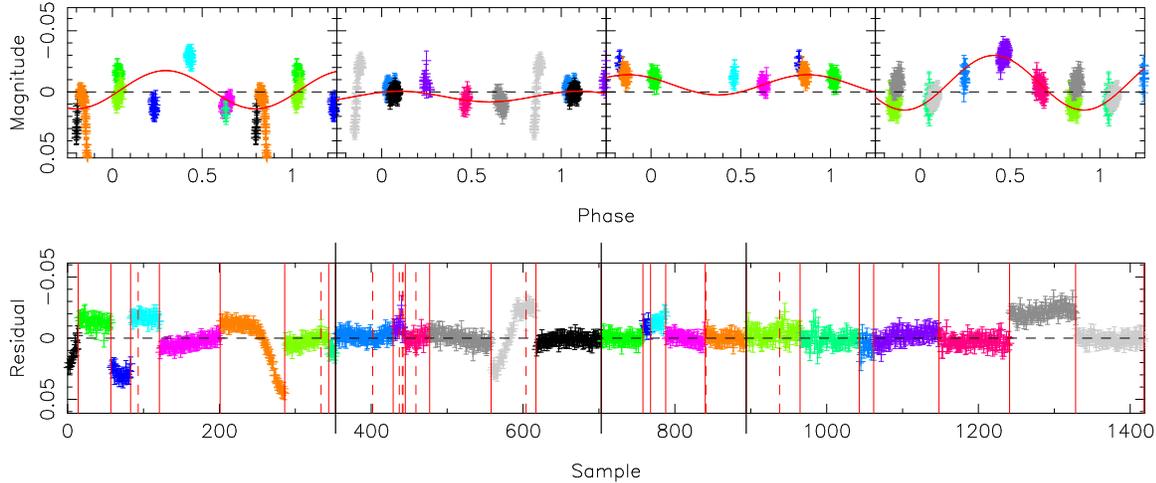}

\caption{Fitting of the INT/WFC $i$-band light curve out-of-eclipse
  variations.  The upper panels show the phase-folded light curve for
  each of the four observing runs, fit with the same period, but a
  different phase, amplitude and zero point in each, with the fit
  overlaid as a solid line.  The lower panel shows the residuals after
  subtracting the fit, plotted as a function of sample number.  The
  dashed horizontal lines show the median light curve level.  The
  vertical lines in the lower plot show the boundaries between nights
  of observations, with the boundaries between observing runs
  (corresponding to the division of the panels in the upper plot)
  denoted by longer black vertical lines.}
\label{290_ooe_fit}
\end{figure*}

The period finding algorithm from \citet{i2006} (based on least
squares fitting of sinusoidal modulations) was modified to fit
different phases, amplitudes and zero-point levels for each of the
four observing runs, fitting for a common period, presumed to be the
rotation period of the star.  The results of this procedure are shown
in Figure \ref{290_ooe_fit}.

Figure \ref{290_ooe_fit} indicates that the fit provides a good
description of the majority of the out-of-eclipse variations, except
in the first five nights and the penultimate night.  This indicates
that a model assuming a constant stellar rotation period, here $4.9\
{\rm days}$ (which is not equal to the binary period or a multiple
thereof), and with moderate evolution of spot coverage over one
month timescales, provides a good description of the majority of the
out-of-eclipse variability.  We note also that \citet{h2002} list a
rotation period of $4.75\ {\rm days}$ for this star, which tends to
confirm our result.

However, Figure \ref{290_ooe_fit} shows that the model we have
presented does not completely describe the variation in the nights
containing eclipses, and this technique is not applicable to the
follow-up data due to the sparse time coverage (providing insufficient
data to fit the rotational modulations in each individual light curve
segment, before the spot configuration changes).

Therefore, we elected to simply reduce the effect of the out of
eclipse variability on the measured eclipse depths by subtracting the
median out of eclipse light curve level from each night containing an
eclipse (see the lower panel of Figure \ref{290_lc}), since this
technique treats the entire composite light curve consistently.  We
caution that this is {\it not} equivalent to modelling the spots
physically, which will be necessary for a more accurate light curve
fit to improve the derived physical parameter estimates.  Fortunately,
in most cases the effect of the spots on the inferred physical
parameters is minor.  For example, in the analysis of the young
eclipsing binary V1174 Ori, \citet{st2004} found that modelling
the out of eclipse variability with spots did not alter the derived
stellar parameters significantly, with changes of $<1$ per cent in the
derived masses and radii, and $<3$ per cent in the ratio of effective
temperatures.

\section{Spectroscopy}
\label{spec_section}

In order to confirm the eclipsing binary nature of the system, and to
derive the orbital parameters, multiple radial velocity measurements
are required.  These were obtained using two instruments: the FLAMES
multi-object fibre-fed \'{e}chelle spectrograph on VLT UT2 in the
optical, and the NOAO Phoenix spectrograph on Gemini South.

\subsection{VLT/FLAMES}

VLT/FLAMES observations were obtained during 2006 November (programme
078.C-0841).  GIRAFFE fibres were allocated to all of our eclipsing
binary candidates in addition to $220$ other suspected ONC members.
A total of five spectra were obtained in each of two standard
settings: HR15n covering $\sim 6450 - 6810\ \angstrom$ with resolving
power $R \sim 17\,000$ and HR21 covering $\sim 8490 - 8980\ \angstrom$
at $R \sim 16\,000$.  Data were reduced using the GIRAFFE Base-Line
Data Reduction Software \citep{blecha2000}, with additional
custom-written software for sky subtraction using $13$ fibres
allocated to blank sky.  The HR15n setting was strongly contaminated
by spatially-variable emission lines from the Orion Nebula, so we
preferred the HR21 setting for determination of radial velocities, and
this has been used henceforth.

Radial velocities were determined by cross-correlation using {\sc
  fxcor} in IRAF\footnote{IRAF is distributed by the National Optical
Astronomy Observatories, which are operated by the Association of
Universities for Research in Astronomy, Inc., under cooperative
agreement with the National Science Foundation.} \citep{tody93}.
Since it is impractical to observe radial velocity standard stars
with FLAMES, we used model atmosphere spectra from the $R \sim
20\,000$ MARCS library \citep{gust2003} to provide the
cross-correlation templates.  Cross-correlations for the present
object were derived using a template with $T_{\rm eff} = 3500\ {\rm
K}$, solar metallicity, and $\log g = 3.5$, in accordance with the
predicted surface gravity from the $1\ {\rm Myr}$ stellar models of
  \citet{bcah98}.  Radial velocity errors were estimated using the
  method of \citet{td79} as implemented in {\sc fxcor}.

\begin{figure}
\centering
\includegraphics[angle=0,width=3.2in,bb=90 240 532 468,clip]{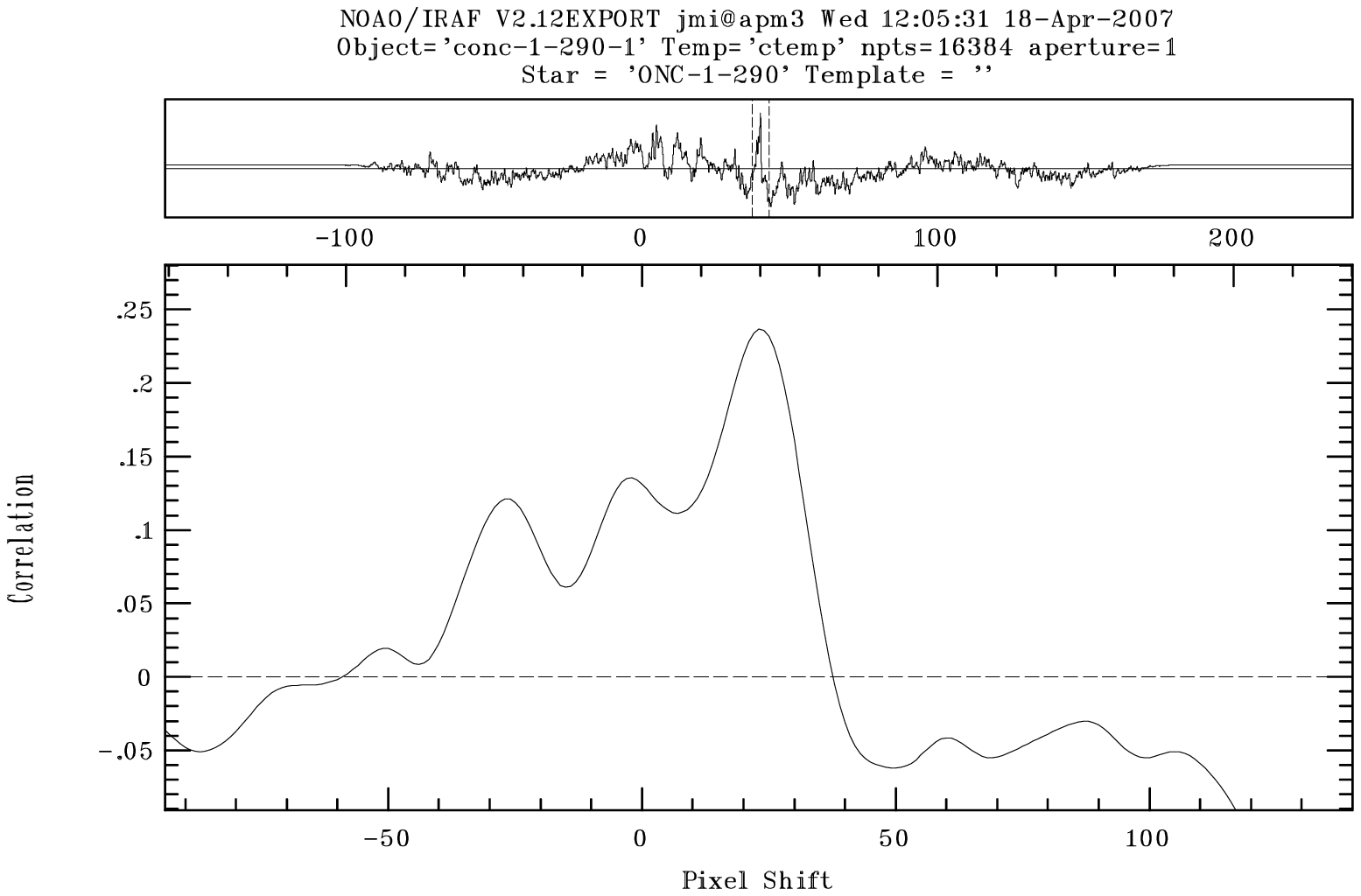}
\includegraphics[angle=0,width=3.2in,bb=90 240 532 468,clip]{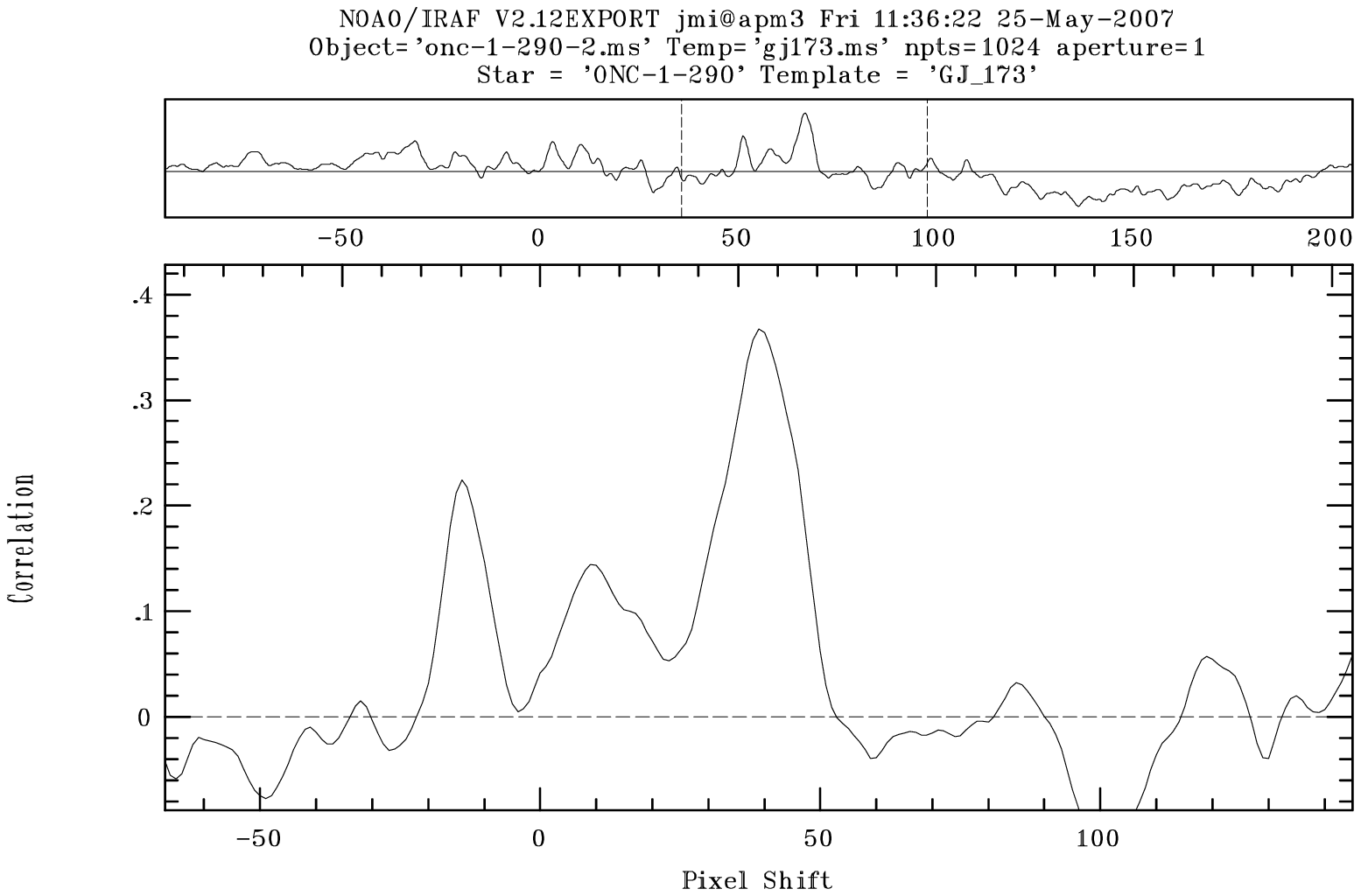}

\caption{Cross-correlation function for the first VLT/FLAMES spectrum
  (top), and a Gemini/Phoenix spectrum for comparison (bottom, see
also \S \ref{phx_sect}), showing a clear triple-lined profile in both
cases.  The central component does not appear to show significant
radial velocity variations.  A three-component Gaussian fit was used
to determine the radial velocities, with cross-correlation peak
heights (left to right) $0.11$, $0.12$, $0.23$ in the VLT/FLAMES
spectrum.}
\label{290_ccf}
\end{figure}

The observed cross-correlation functions show clear triple-lined
profiles (see Figure \ref{290_ccf}) in a total of $7$ epochs
(including all the radial velocity data, see also \S \ref{phx_sect})
around maximum radial velocity separation, with the two outer
components (the primary and secondary stars in the eclipsing binary)
exhibiting radial velocity variations.  The central component
(hereafter the tertiary -- although note that this star is not
necessarily physically associated with the binary) does not appear to
show radial velocity variations.

\begin{figure}
\centering
\includegraphics[angle=270,width=3.2in]{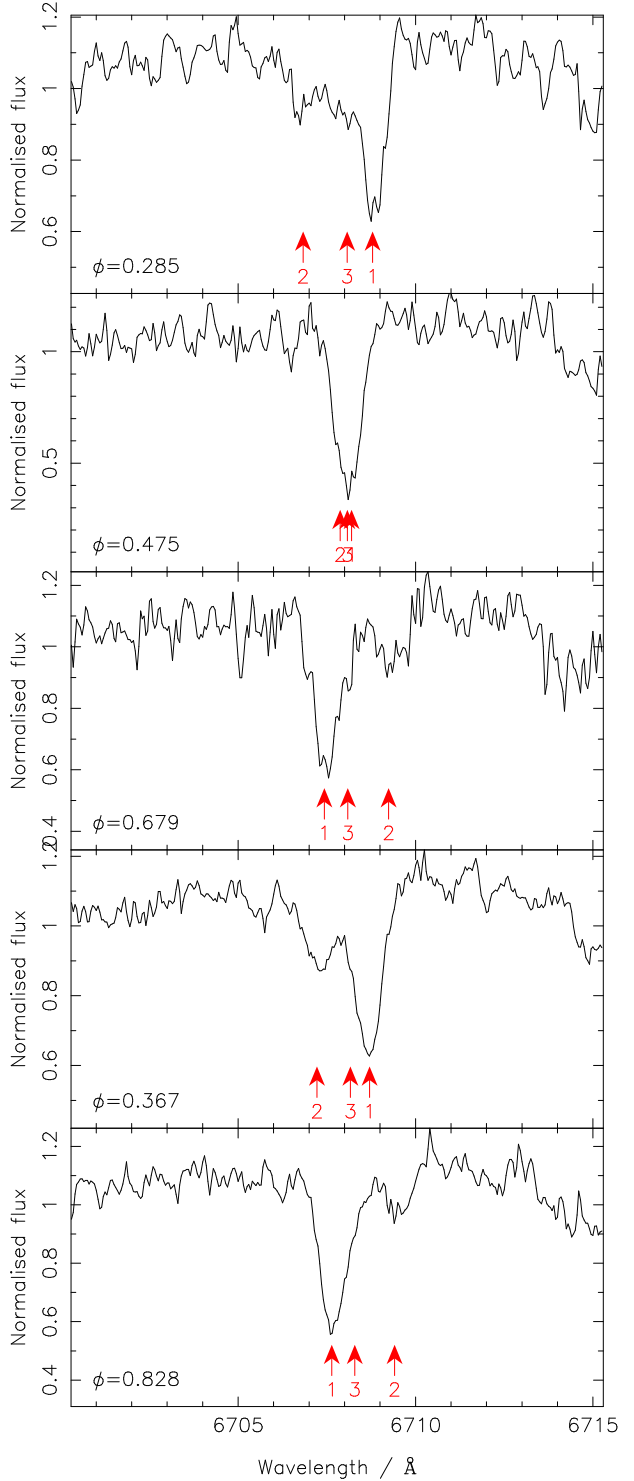}

\caption{Lithium $6707.8\ \angstrom$ line profiles for the five
  VLT/FLAMES spectra in the HR15n setting.  Arrows show the predicted
  positions of the lines from the three components (1, 2 and 3 are
  primary, secondary and tertiary respectively), derived from the
  radial velocities.  The primary and secondary show clear Lithium
  absorption, and the tertiary may also show it, but this line, if
  present, is not well-resolved from the primary in any of the spectra.}
\label{290_li}
\end{figure}

We note that strong Li{\sc i} $6707.8\ \angstrom$ absorption is clearly
visible in the HR15n setting, which indicates youth, and hence
membership of the ONC.  The Lithium line profiles from our $5$ HR15n
spectra are shown in Figure \ref{290_li}.  Equivalent widths were
estimated by fitting three-component Gaussian models to the two
best-resolved epochs (the first and third in Fig. \ref{290_li}), at
the expected wavelengths from the radial velocities for each
component.  The values obtained were $0.32 \pm 0.05\ \angstrom$ for
the primary, $0.18 \pm 0.04\ \angstrom$ for the secondary, and $0.12
\pm 0.05\ \angstrom$ for the tertiary (errors estimated using
bootstrapping).  Note that in order to compare to the values for
single stars, these measured equivalent widths must be corrected for
the relative luminosities of the stars.

Unfortunately, we are unable to correct the Li{\sc i} equivalent
widths for spectral veiling at present, since we do not see suitable
features in our spectra (e.g. Ni{\sc i} $6643\ \angstrom$, Fe{\sc i}
$6663\ \angstrom$ or V{\sc i} $6625\ \angstrom$ as used by
\citealt{palla2007}) with which to compute the veiling, despite
the majority of our spectra having very good signal to noise ratios
($\sim 40\ {\rm pix^{-1}}$ or $180\ {\rm \AA^{-1}}$).  The values we
have computed must therefore be considered to give lower limits on the
measured Lithium abundance.  Nevertheless, assuming the primary to
contribute $1/2$ of the light in the system, the implied equivalent
width is $\sim 0.6\ \angstrom$, which is a typical value for ONC stars
of this mass (e.g. see \citealt{sa2005}). 

\subsection{Gemini/Phoenix}
\label{phx_sect}

Additional radial velocity measurements were obtained using the
near-infrared spectrograph Phoenix \citep{hinkle2002} on Gemini South,
on three nights: 2006 December 1, 2 and 4 (programme GS-2006B-C-7).
Following \citet{maz2002}, we used the H6420 order sorting filter and
the widest ($4\ {\rm pix} \approx 0.34\ {\rm arcsec}$) slit, with a
central wavelength of $1.558\ {\rm \mu m}$ to give a wavelength range
of $1.554 - 1.562\ {\rm \mu m}$ at resolving power $R \sim 35\,000$.
Exposures were taken in {\it AB} pairs, nodding along the slit, to aid
subtraction of the sky background and residual dark current features
in the detector.  A total of eight epochs were obtained for this
object, with average signal to noise ratios of $\sim 60\ {\rm
  pix^{-1}}$ or $220\ {\rm A^{-1}}$.  Data were reduced using standard
{\sc IRAF} long-slit reduction procedures, and cross-correlations with
the radial velocity standard star GJ 173 (M1.5 spectral type) in {\sc
  fxcor} were used to derive radial velocities, since this template
gave the largest cross-correlation signal.  This suggests that the
primary spectral type is close to M1.5.  Using the effective
temperature scale of \citet{ck79}, this corresponds to $T_{\rm eff} =
3590\ {\rm K}$.

Despite the high resolution and signal to noise ratio of our data, it
is difficult to see the individual line profiles in the Phoenix
spectra.  This may be partly due to the very strong dependence on
spectral type in this wavelength region (e.g. see \citealt{bender05}),
where the atomic lines become weak moving to the later spectral type
secondary and tertiary components.  This also partly explains the 
reduction in cross-correlation signal for these objects (see Figure
\ref{290_ccf}), since we used an earlier-type template with strong
atomic lines.

We note that the unavoidable use of different templates for the two
instruments may introduce systematic offsets between the two sets of
radial velocities.  We have tried to minimise this source of error by
using templates with similar effective temperatures.

\subsection{Spectroscopic orbit solution}

The photometry and initial VLT/FLAMES radial velocities were found to
be consistent with the photometrically-derived period, but the
Gemini/Phoenix velocities were not.  A consistent solution was found
at twice the initial period, or $5.2991\ {\rm days}$.  Re-examination
of the light curve indicated that three secondary eclipses had in fact
been observed, and that these had depths of $\sim 0.03\ {\rm mag}$ in
$i$-band.  This is simply a result of the improved phase coverage
obtained by including the Gemini/Phoenix radial velocity data.

\begin{figure}
\centering
\includegraphics[angle=270,width=3.2in]{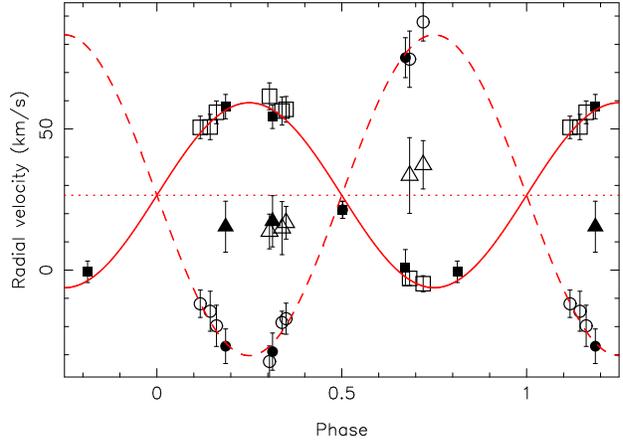}

\caption{Radial velocity curve, folded on a period of $5.2991\ {\rm
    days}$, where phases $0, 1, \ldots$ are defined to be at mid
    secondary eclipse.  The curves show the fit for a circular
    orbit, to the primary (solid line) and secondary (dashed line)
    radial velocities.  The dotted line indicates the systemic radial
    velocity.  Points with error bars show the radial velocity
    measurements for all three components of the system (primary:
    squares, secondary: circles, tertiary: triangles), filled symbols
    denote radial velocity points from VLT/FLAMES, and open symbols
    points from Gemini/Phoenix.  Velocities for the system components
    were only measured when the corresponding peaks were visible in
    the cross correlation functions, so some data points are missing
    for the secondary and tertiary components, particularly close to
    phases $0$ and $0.5$.  The tertiary velocities are poorly measured
    and are strongly influenced by blending of the cross-correlation
    profiles (e.g. see Figure \ref{290_ccf}), so the apparent
    variations for this component are not considered significant in
    the present data.}
\label{290_rv}
\end{figure}

Figure \ref{290_rv} shows the resulting phase-folded radial velocity
curve, with period and phase zero-point fixed at the
photometrically-determined values.  Neither the light curve nor radial
velocity curve appear to show significant deviation from the
predictions for a circular orbit, with a formal fit giving
eccentricity $e = 0.004 \pm 0.036$, so we have assumed zero
eccentricity henceforth.

Table \ref{290_rvparm} gives the orbital parameters
derived from the radial velocity fit.

\begin{table}
\centering

\begin{tabular}{lll}
\hline
\hline
Systemic velocity        &$v_{\rm sys}$  &$26.55 \pm 0.87\ \kms$ \\
Primary semi-amplitude   &$K_1$          &$32.8 \pm 1.3\ \kms$ \\
Secondary semi-amplitude &$K_2$          &$56.9 \pm 2.2\ \kms$ \\
\hline
Mass ratio ($M_2/M_1$) &$q$            &$0.577 \pm 0.032$ \\
Semimajor axis         &$a \sin i$     &$9.38 \pm 0.27\ \rsun$ \\
Total mass             &$M \sin^3 i$   &$0.396 \pm 0.019\ \msun$ \\
Primary mass           &$M_1 \sin^3 i$ &$0.251 \pm 0.023\ \msun$ \\
Secondary mass         &$M_2 \sin^3 i$ &$0.145 \pm 0.012\ \msun$ \\
\hline
Reduced $\chi^2$       &$\chi^2_\nu$   &0.60 \\
\hline
\hline
\end{tabular}

\caption{System parameters derived from the radial velocity curve fit
  in Figure \ref{290_rv}.}
\label{290_rvparm}
\end{table}

The measured systemic velocity ($v_{\rm sys}$) is $26.55 \pm 0.87\
\kms$.  This is very close to the systemic velocity of the ONC, of
$25 \pm 1.5\ \kms$ (\citealt{smmv99}; \citealt{sa2005}), which
provides a further kinematic confirmation of cluster membership.

Since the tertiary lies approximately at the ONC systemic velocity in
the cross-correlations, it seems likely that this star is also a
member of the ONC.  If this is the case, the measured stellar density
in the ONC (e.g. \citealt{h97} found $\sim 1600$ members with $I <
17.5$ over a $0.2\ {\rm sq. deg}$ region) indicates a probability of
$\sim 10^{-4}$ of a chance superposition of another ONC member with
the binary, within $0.5''$ (the best seeing experienced in our
observing runs).  It thus seems likely that the system is a physical
triple.

\section{Light curve analysis}
\label{lcanal_section}

The $i$-band light curve was analysed using {\sc jktebop}
(\citealt*{south2004a}; \citealt{south2004b}), a modified version of
{\sc ebop} (Eclipsing Binary Orbit Program; \citealt{pe81};
\citealt{e81}).  These codes use a model in which the discs of the
stars are approximated using biaxial ellipsoids \citep{nd72}.  This
approximation is only applicable in well-detached systems where the
stars are close to spherical, as in the present case.  Note that star
spot modelling and handling of multi-band light curves are not
implemented in these codes, and must be done externally, as we have in
the present work.  The {\sc jktebop} code contains several
enhancements, including fitting of the sum and ratio of the stellar
radii, the use of the Levenberg-Marquardt minimization algorithm for
least-squares fitting \citep{press93}, and of particular interest for
the present discussion, Monte Carlo simulation to determine robust
error estimates for the stellar parameters, which has been shown to
give  reliable results \citep{south2005}.  Note that we have not used
the $V$-band light curve in the present analysis, since there are no
secondary eclipses observed, and the photometric precision reached is
somewhat poorer.  The observed primary eclipses have very similar
depth to the $i$-band eclipses.

The light curve analysis for the present system is complicated by the
known presence of third light.  From Figure \ref{290_ccf}, it appears
that the cross-correlation peaks for the secondary and tertiary are of
very similar height and shape, which suggests that the third light
star has a similar flux and spectral type to the secondary.  We
therefore expect it to contribute $\sim 1/4$ of the light in the
system, assuming a $2:1:1$ ratio of $i$-band luminosity.

Due to the need to fit the third light contribution, we fixed as many
parameters as possible, to improve stability.  In $I$-band, limb
darkening is of much lower importance than for bluer passbands, so we
fixed these parameters.  We used a linear limb darkening law:
\begin{equation}
{I_\lambda(\mu)\over{I_\lambda(1)}} = 1 - x_\lambda \left(1 - \mu\right)
\label{limbdark_eq}
\end{equation}
where $\mu = \cos \theta$ is the cosine of the angle between the
line-of-sight and the normal to the stellar surface, $I_\lambda(\mu)$
is the observed surface intensity at wavelength $\lambda$ as a
function of $\mu$, and $x_\lambda$ is the linear limb darkening
coefficient.  Values of $x_\lambda$ of $0.7208$ for the primary star
($T_{\rm eff} = 3300\ {\rm K}$, $\log g = 4.0$, $[M/H] = 0$) and
$0.7341$ ($T_{\rm eff} = 3100\ {\rm K}$, $\log g = 4.0$, $[M/H] = 0$)
for the secondary star were adopted from \citet{claret2004}, for the
SDSS $i'$ passband, which is the closest match to the SDSS-like $i$
filters used in this work.  The gravity darkening exponents were also
fixed, at $\beta = 0.08$, a value appropriate for stars with
convective envelopes \citep{lucy67}, and the option in {\sc jktebop}
to calculate the reflection effect was used, rather than fitting for
it.  This is generally a reasonable approach for well-detached
systems.

The remaining parameters were allowed to vary.  These are the sum
of the radii, $r_1+r_2$ (where $r_i = R_i / a$, the radius divided by
semimajor axis, a parameter which can be determined from the light
curve analysis alone), the radius ratio $k = r_2 / r_1$, orbital
inclination $i$, surface brightness ratio $J$ (again defined as
secondary divided by primary such that $J \le 1$) and fractional
luminosity of the third light $L_3$.  We also allowed the period $P$
and phase zero-point $T_0$ (defined as the heliocentric Julian day of
the secondary eclipses) to vary, to refine the existing values.  Note
that the quantity $J$ and the luminosities quoted here are for
$i$-band.

Note that the {\sc ebop} codes model only single-band light curves,
and do not use model atmospheres, so it is not necessary to assume an
effective temperature or surface gravity in the fit itself: these
parameters enter only through the (assumed) limb darkening
coefficients, and we expect this dependence to be weak in $i$-band.

\begin{figure}
\centering
\includegraphics[angle=270,width=3.2in]{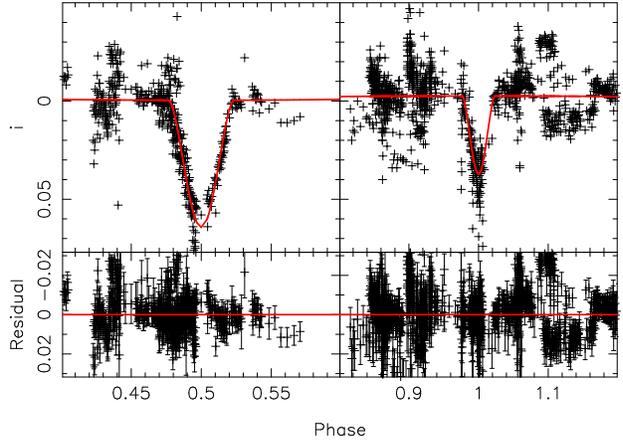}

\caption{Phase-folded combined $i$-band light curve, with phase $0$
  defined to be at mid-secondary eclipse, as Figure \ref{290_rv}.  The
  upper panel shows the light curve, with the fit overlaid (solid line), and
  the lower panel shows the residuals (data - model).  The panels show
  a magnified view of the regions around primary eclipse (left)
  and secondary eclipse (right).  The scatter in the secondary
  eclipse is larger than the primary since this was only observed in
  the follow-up observations and not the INT/WFC survey photometry
  (which was of better photometric precision).}
\label{290_i_fit}
\end{figure}

\begin{figure*}
\centering
\includegraphics[angle=270,width=6in]{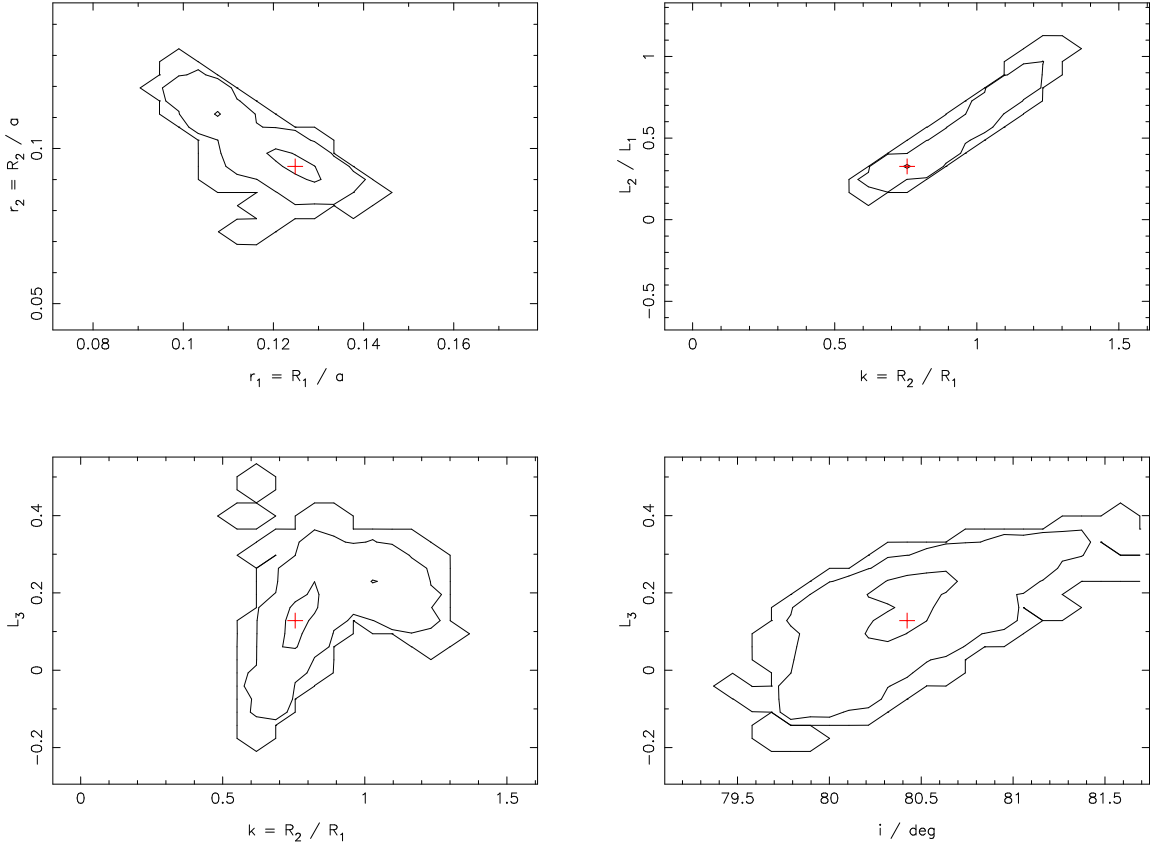}

\caption{Confidence intervals on the light curve fit parameters from
  the Monte Carlo simulations.  The contours show the $68.3$, $95.4$
  and $99.73$ per cent confidence levels, corresponding to $1$, $2$
  and $3\sigma$ standard errors.  The crosses indicate the
  best-fitting values.  The panels show: $r_2$ versus $r_1$ (top
  left), luminosity ratio $L_2 / L_1$ versus radius ratio $k$,
  fractional third light luminosity $L_3$ versus $k$, and $L_3$ versus
  orbital inclination $i$.  Degeneracies between $r_1$ and $r_2$, and
  $L_2 / L_1$ and $k$ (due to the existence of solutions for $k <
  1$ and for $k > 1$ as discussed in the text) are clearly visible.
  Note that the $3 \sigma$ contours are likely to be underestimated
  due to the limited number of Monte Carlo samples used.}
\label{290_contours}
\end{figure*}

Figure \ref{290_i_fit} show the $i$-band phase-folded light curve,
with the fit overlaid.  The system parameters derived from the light
curve fitting are given in Table \ref{290_lcparm}.  Errors were
estimated using the Monte Carlo analysis, with $10\,000$ iterations,
the results of which are also shown in Figure \ref{290_contours}.

\begin{table}
\centering

\begin{tabular}{lll}
\hline
\hline
Radius sum                &$(R_1 + R_2)/a$ &$0.2191^{+0.0035}_{-0.0039}$ \\
Radius ratio              &$k$             &$0.75^{+0.30}_{-0.04}$ \\
Orbital inclination       &$i$             &$80.42^{+0.27}_{-0.26}\ {\rm degrees}$ \\
Surface brightness ratio  &$J$             &$0.573^{+0.045}_{-0.011}$ \\
Third light ratio         &$L_3$           &$0.128^{+0.111}_{-0.058}$ \\
Orbital Period            &$P$             &$5.299180^{+0.000013}_{-0.000014}\ {\rm days}$ \\
Phase zeropoint (HJD)     &$T_0$           &$2449704.45279^{+0.00998}_{-0.00946}$ \\
\hline
Primary radius            &$R_1/a$         &$0.1248^{+0.0031}_{-0.0184}$ \\
Secondary radius          &$R_2/a$         &$0.0942^{+0.0182}_{-0.0029}$ \\
Luminosity ratio          &$L_2 / L_1$     &$0.3265^{+0.3635}_{-0.0372}$ \\
Temperature ratio         &$T_2 / T_1$     &$0.8700^{+0.0168}_{-0.0041}$ \\
\hline
\hline
\end{tabular}

\caption{System parameters derived from the light curve fit
  in Figure \ref{290_i_fit}.  Note that $J$, $L_3$ and $L_2 / L_1$ are
  the quantities for $i$-band.  $68$ per cent confidence intervals are
  quoted.  Note that since the effective temperatures are unknown, the
  temperature ratio was calculated assuming equal bolometric
  corrections for both components.}
\label{290_lcparm}
\end{table}

The Monte Carlo algorithm uses the best fit to generate a synthetic
light curve, injecting Gaussian noise with amplitude determined by the 
observational errors to produce a simulated light curve, which is
then fit to determine a new set of parameters.  The errors can be
estimated using the distribution of the parameters from a large
number of realisations of this process.  See also \citet{south2004a}
and \citet{south2004b} for more details.

Since the luminosity ratio was not constrained in the analysis, the
solution with the radii of the two stars swapped ($k > 1$ or $R_2 >
R_1$) is also formally permitted by the fit, which leads to highly
asymmetric errors for $k$, $L_2/L_1$ and the two fractional radii.

Table \ref{290_parm} summarises the final system parameters, combining
the radial velocity and light curve information.  Note the extremely
large stellar radii and low surface gravities, as expected for a very
young system.

\begin{table}
\centering

\begin{tabular}{lll}
\hline
\hline
Primary mass              &$M_1$           &$0.262^{+0.025}_{-0.024}\ \msun$ \\
Secondary mass            &$M_2$           &$0.151^{+0.013}_{-0.013}\ \msun$ \\
Primary radius            &$R_1$           &$1.189^{+0.039}_{-0.175}\ \rsun$ \\
Secondary radius          &$R_2$           &$0.897^{+0.170}_{-0.034}\ \rsun$ \\
\hline
Semimajor axis            &$a$             &$9.52^{+0.27}_{-0.27}\ \rsun$ \\
Primary gravity           &$\log(g_1)$     &$3.706^{+0.137}_{-0.025}\ {\rm cm\ s^{-1}}$ \\
Secondary gravity         &$\log(g_2)$     &$3.711^{+0.029}_{-0.155}\ {\rm cm\ s^{-1}}$ \\
\hline
\hline
\end{tabular}

\caption{Physical parameters derived from the combined radial velocity
  and light curve fitting.  Note that the errors on $M_1$ and $M_2$,
  and $R_1$ and $R_2$ are not independent since they are calculated
  from the mass ratio and radius ratio.}
\label{290_parm}
\end{table}

We caution that the effects of the out-of-eclipse modulations are
non-negligible in this system, and these have not been modelled at
present due to the lack of suitable light curve data.  Therefore, the
parameters derived from the light curve fitting may be affected by
systematic errors due to not properly accounting for the spot-induced
photometric variations.  Improved light curves will be required to
resolve this problem.

\section{Comparison with stellar models}
\label{modelcomp_section}

We are unable to determine accurate effective temperatures at the
present time, so the only comparison which can be made to stellar
models with the present data-set is in the mass-radius plane.
Figure \ref{mrrel} shows a comparison of the observations for the
present system with the Lyon group stellar models.  The primary and
secondary are consistent within the $1 \sigma$ error bars with the $2\
{\rm Myr}$ model of \citet{bcah98}, but $\sim 2\sigma$ away from the
$1\ {\rm Myr}$ model.  The primary is also consistent within $1
\sigma$ with the $3\ {\rm Myr}$ model.  The value of $2\ {\rm Myr}$ is
fully consistent with the upper end of the canonical $1 \pm 1\ {\rm
  Myr}$ age for the ONC \citep{h97}.

\begin{figure}
\centering
\includegraphics[angle=270,width=3.2in]{mrrel.ps}

\caption{Mass-radius relation for low-mass stars and eclipsing
  binaries.  The present system is shown by the black points with
  error bars, and the lines show pre-main-sequence NextGen ($\alpha =
  1.0$, solid lines; \citealt{bcah98}) and DUSTY (dashed lines;
  \citealt{cbah2000}) models from the Lyon group, at five ages (top to
  bottom): $1\ {\rm Myr}$ (grey line), $2\ {\rm Myr}$ (blue), $3\ {\rm
  Myr}$ (red), $10\ {\rm Myr}$ (magenta) and $1\ {\rm Gyr}$ (black).
  Systems shown as red points with error bars are existing pre-main
  sequence binaries from \citet{cov2001}, \citet{cov2004},
  \citet{st2004}, \citet{hebb2006} and \citet{st2007}.  The small grey
  points with error bars are a compilation of field systems from
  \citet{del2000}, \citet{lane2001}, \citet{seg2003}, \citet{lm2004},
  \citet{bou2005}, \citet{pont2005} and \citet{lm2006}.  We have opted
  to show results from the literature to produce a figure summarising
  the present empirical constraints on the PMS mass-radius relation.
  The DUSTY models are included to show the predicted behaviour in the
  brown dwarf domain.}
\label{mrrel}
\end{figure}

\begin{figure*}
\centering
\includegraphics[angle=270,width=6in]{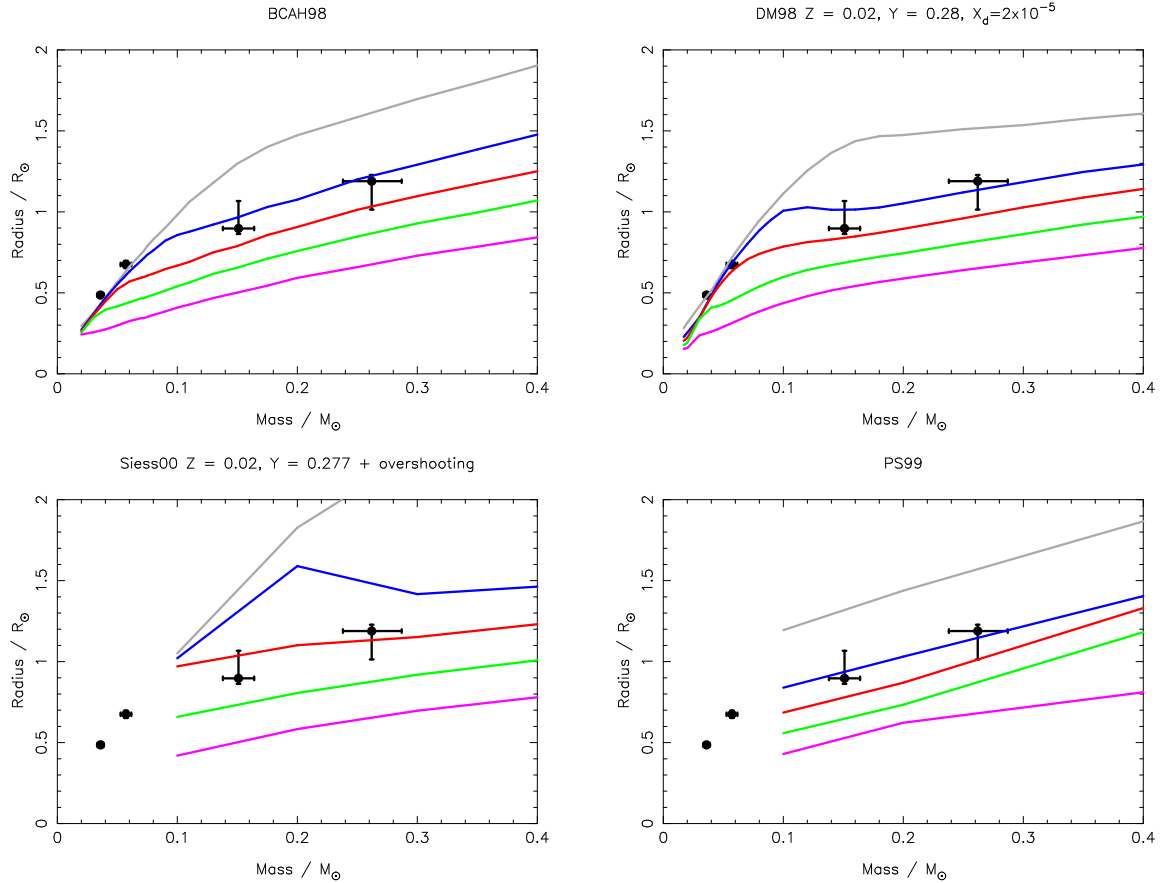}

\caption{Comparison of the eclipsing binary system with several models
  of low-mass stars.  The EB data are shown by the black points with
  error bars, and the lines show pre-main sequence models for (top to
  bottom): $1\ {\rm Myr}$ (grey line), $2\ {\rm Myr}$ (blue), $3\ {\rm
  Myr}$ (red), $5\ {\rm Myr}$ (green) and $10\ {\rm Myr}$ (magenta).
  Top left: \citet{bcah98} for solar metallicity and $\alpha = 1.0$,
  top right: \citet{dm98} for $Z = 0.02$, $Y = 0.28$, $X_d =
  2\times10^{-5}$, bottom left: \citet{siess2000} for $Z = 0.02$, $Y =
  0.277$, with (moderate) convective overshooting, bottom right:
  \citet{ps99}.  The models have been interpolated where necessary to
  provide points at the required ages.  The brown dwarf eclipsing
  binary system of \citet{st2006} (using the parameters from
  \citealt{st2007}) is shown to allow the inferred ages to be
  compared, since this is suspected to also be a member of the ONC,
  and should therefore be assigned a similar age to JW 380.  Note that
  the \citet{siess2000} and \citet{ps99} models do not provide points
  below $0.1\ \msun$.}
\label{modelcomp}
\end{figure*}

Figure \ref{modelcomp} shows the position of the eclipsing binary
relative to models from \citet{bcah98}, \citet{dm98},
\citet{siess2000} and \citet{ps99}.  The results for the
\citet{bcah98}, \citet{dm98} and \citet{ps99} are consistent with the
$2\ {\rm Myr}$ inferred age from above; although there are clearly
visible differences between the results of these sets of tracks
(especially when comparing \citealt{dm98} to the others), the
measurement errors are too large to distinguish them at  present.
Comparing the data with the models of \citet{siess2000} would indicate
a slightly older age of $\sim 3\ {\rm  Myr}$.  However, there is a
visible discontinuity in the latter tracks at $2\ {\rm Myr}$, and they
are not well-sampled over this mass range.

Given the present observational errors, it is not yet meaningful to
perform a more detailed comparison with stellar models.  We plan to
obtain improved photometry and radial velocity measurements to resolve
this, in addition to spectroscopy covering suitable lines for
determining effective temperatures and flux ratios.

\section{Conclusions}
\label{concl_section}

We have reported the detection of a new pre-main-sequence eclipsing
binary, JW 380, which appears to be a member of the Orion Nebula
Cluster (membership probability $0.99$, systemic velocity within $1
\sigma$ of the cluster systemic velocity).  The system consists of a
$0.26 \pm 0.02$, $0.15 \pm 0.01\ \msun$ eclipsing binary with period
$5.299180 \pm 0.000013\ {\rm days}$, blended with a significant
($\sim 20$ per cent of the total luminosity) third light component,
which appears to show no significant radial velocity variations over
the (limited) observing window.

Modelling the combined light and radial velocity curves for the system
yielded stellar radii of $1.19^{+0.04}_{-0.18}\ \rsun$ and
$0.90^{+0.17}_{-0.03}\ \rsun$ for the primary and secondary of the
binary system.  These large radii are consistent with the $2\ {\rm
  Myr}$ age stellar models of \citet{bcah98}, \citet{dm98} and
\citet{ps99}, and the $3\ {\rm Myr}$ age models of \citet{siess2000}
given the masses of the system components.  The systemic radial
velocity is consistent with that of the ONC, and the detection of a
clear Li{\sc i} $6707.8\ \angstrom$ absorption feature is also
suggestive of youth, and hence ONC membership.

\section*{Acknowledgments}

The Isaac Newton Telescope is operated on the island of La Palma by
the Isaac Newton Group in the Spanish Observatorio del Roque de los
Muchachos of the Instituto de Astrofisica de Canarias.  Based in part
on observations collected at the European Southern Observatory, Chile,
as part of ESO programme 078.C-0841.  Also based on observations
obtained with the Phoenix infrared spectrograph, developed and
operated by the National Optical Astronomy Observatory, at the
Gemini Observatory, which is operated by the Association of
Universities for Research in Astronomy, Inc., under a cooperative
agreement with the NSF on behalf of the Gemini partnership: the
National Science Foundation (United States), the Particle Physics and
Astronomy Research Council (United Kingdom), the National Research
Council (Canada), CONICYT (Chile), the Australian Research Council
(Australia), CNPq (Brazil) and CONICET (Argentina).  Based on
observations obtained at Cerro Tololo Inter-American Observatory, a
division of the National Optical Astronomy Observatories, which is
operated by the Association of Universities for Research in Astronomy,
Inc. under cooperative agreement with the National Science Foundation.
This publication makes use of data products from the Two Micron All
Sky Survey, which is a joint project of the University of
Massachusetts and the Infrared Processing and Analysis
Center/California Institute of Technology, funded by the National
Aeronautics and Space Administration and the National Science
Foundation.  This research has also made use of the SIMBAD database,
operated at CDS, Strasbourg, France.

JI gratefully acknowledges the support of a PPARC studentship, and SA
the support of a PPARC postdoctoral fellowship.  This work was supported
by NSF Career grant AST-0349075, and by a Cottrell Scholar award from
the Research Corporation, to K.G.S.  We would like to thank Francesco
Palla for supplying model tracks used in \S \ref{modelcomp_section},
John Southworth for the {\sc jktebop} code, Gwendolyn Meeus for
VLT/ISAAC spectroscopy, Thierry Morel, Hans van Winckel and Conny
Aerts for data from the $1.2\ {\rm m}$ Mercator telescope, and the
anonymous referee for feedback which has helped to improve the paper.


\begin{thebibliography}{}
\bibitem[\protect\citeauthoryear{Aigrain \& Irwin}{2004}]{ai2004}
  Aigrain S., Irwin M., 2004, MNRAS, 350, 331
\bibitem[\protect\citeauthoryear{Aigrain et al.}{2007}]{a2007}
  Aigrain S., Hodgkin S., Irwin J., Hebb L., Irwin M., Favata F.,
  Moraux E., Pont F., 2007, MNRAS, 375, 29
\bibitem[\protect\citeauthoryear{Baraffe et al.}{1998}]{bcah98}
  Baraffe I., Chabrier G., Allard F., Hauschildt P.H., 1998, A\&A,
  337, 403
\bibitem[\protect\citeauthoryear{Baraffe et al.}{2002}]{bcah2002}
  Baraffe I., Chabrier G., Allard F., Hauschildt P.H., 2002, A\&A,
  382, 563
\bibitem[\protect\citeauthoryear{Bender et al.}{2005}]{bender05}
  Bender C., Simon M., Prato L., Mazeh T., Zucker S., 2005, AJ, 129,
  402
\bibitem[\protect\citeauthoryear{Blecha et al.}{2000}]{blecha2000}
  Blecha A., Cayatte V., North P., Royer F., Simond G., 2000, in
  Proc. SPIE vol 408, p. 467-474, Optical and IR Telescope
  Instrumentation and Detectors, Eds. M. Iye, A.F. Moorwood
\bibitem[\protect\citeauthoryear{Bouchy et al.}{2005}]{bou2005}
  Bouchy F., Pont F., Melo C., Santos N.C., Mayor M., Queloz D., Udry
  S., 2005, A\&A, 431, 1105
\bibitem[\protect\citeauthoryear{Burrows et al.}{1997}]{burrows97}
  Burrows A. et al., 1997, ApJ, 491, 856
\bibitem[\protect\citeauthoryear{Chabrier \& Baraffe}{1997}]{cb97}
  Chabrier G., Baraffe I., 1997, A\&A, 327, 1039
\bibitem[\protect\citeauthoryear{Chabrier et al.}{2000}]{cbah2000}
  Chabrier G., Baraffe I., Allard F., Hauschildt P., 2000, ApJ, 542, 464
\bibitem[\protect\citeauthoryear{Claret}{2004}]{claret2004}
  Claret A., 2004, A\&A, 428, 1001
\bibitem[\protect\citeauthoryear{Cohen \& Kuhi}{1979}]{ck79}
  Cohen M., Kuhi L.V., 1979, ApJS, 41, 743
\bibitem[\protect\citeauthoryear{Covino et al.}{2001}]{cov2001}
  Covino E., Melo C., Alcal\'{a} J.M., Torres G., Fern\'{a}ndez M.,
  Frasca A., Paladino R., 2001, A\&A, 375, 130
\bibitem[\protect\citeauthoryear{Covino et al.}{2004}]{cov2004}
  Covino E., Frasca A., Alcal\'{a} J.M., Paladino R., Sterzik M.F.,
  2004, A\&A, 427, 637
\bibitem[\protect\citeauthoryear{Dahm \& Hillenbrand}{2007}]{dh07}
  Dahm S.E., Hillenbrand L., 2007, AJ, 133, 2072
\bibitem[\protect\citeauthoryear{D'Antona \& Mazzitelli}{1997}]{dm97}
  D'Antona F., Mazzitelli I., 1997, Mem. Soc. Astron. Ital., 68, 807
\bibitem[\protect\citeauthoryear{D'Antona \& Mazzitelli}{1998}]{dm98}
  D'Antona F., Mazzitelli I., 1998, ASP Conf. Series 134, p. 442,
  Brown dwarfs and extrasolar planets, eds. R. Rebolo, E.L. Martin,
  M.R. Zapatero Osorio
\bibitem[\protect\citeauthoryear{Delfosse et al.}{2000}]{del2000}
  Delfosse X., Forveille T., S\'{e}gransan D., Beuzit J.-L., Udry S.,
  Perrier C., Mayor M., 2000, A\&A, 364, 217
\bibitem[\protect\citeauthoryear{Etzel}{1981}]{e81}
  Etzel P.B., 1981, in Carling E. B., Kopal Z., eds, Photometric and
  Spectroscopic Binary Systems, NATO ASI Ser. C., 69. Kluwer,
  Dordrecht, p. 111
\bibitem[\protect\citeauthoryear{Getman et al.}{2005}]{getman2005}
  Getman K.V., et al., 2005, ApJS, 160, 319
\bibitem[\protect\citeauthoryear{Girardi et al.}{2000}]{gir2000}
  Girardi L., Bressan A., Bertelli G., Chiosi C., 2000, A\&AS, 141, 371
\bibitem[\protect\citeauthoryear{Gustafsson et al.}{2003}]{gust2003}
  Gustafsson B., Edvardsson B., Eriksson K., Mizuno-Wiedner M.,
  J{\o}rgensen U.G., Plez B., 2003, in ASP Conf. Series vol 288, p. 331,
  Stellar Atmosphere Modeling, Eds. I. Hubeny, D. Mihalas, K. Werner
\bibitem[\protect\citeauthoryear{Hebb et al.}{2006}]{hebb2006}
  Hebb L., Wyse R.F.G., Gilmore G., Holtzman J., 2006, AJ, 131, 555
\bibitem[\protect\citeauthoryear{Herbst et al.}{2002}]{h2002}
  Herbst W., Bailer-Jones C.A.L., Mundt R., Meisenheimer K.,
  Wackermann R., 2002, A\&A, 396, 513
\bibitem[\protect\citeauthoryear{Hillenbrand}{1997}]{h97}
  Hillenbrand L., 1997, AJ, 113, 1733
\bibitem[\protect\citeauthoryear{Hinkle et al.}{2002}]{hinkle2002}
  Hinkle K.H., et al., 2002, Proc. SPIE 3354, 810
\bibitem[\protect\citeauthoryear{Hodgkin et al.}{2006}]{hodg06}
  Hodgkin S.T., Irwin J.M., Aigrain S., Hebb L., Moraux E., Irwin
  M.J., 2006, AN, 327, 9
\bibitem[\protect\citeauthoryear{Honeycutt}{1992}]{h92}
  Honeycutt R.K., 1992, PASP, 104, 435
\bibitem[\protect\citeauthoryear{Irwin et al.}{2006}]{i2006}
  Irwin J., Aigrain S., Hodgkin S., Irwin M., Bouvier J., Clarke C.,
  Hebb L., Moraux E., 2006, MNRAS, 370, 954
\bibitem[\protect\citeauthoryear{Irwin et al.}{2007}]{i2007}
  Irwin J., Irwin M., Aigrain S., Hodgkin S., Hebb L., Moraux E.,
  2007, MNRAS, 375, 1449
\bibitem[\protect\citeauthoryear{Jones \& Walker}{1988}]{jw88}
  Jones B.F., Walker M.F., 1988, AJ, 95, 1755
\bibitem[\protect\citeauthoryear{Lane et al.}{2001}]{lane2001}
  Lane B.F., Boden A.F., Kulkarni S.R., 2001, ApJ, 551, 81
\bibitem[\protect\citeauthoryear{Leggett}{1992}]{leg92}
  Leggett S.K., 1992, ApJS, 82, 351
\bibitem[\protect\citeauthoryear{Lopez-Morales}{2004}]{lm2004}
  Lopez-Morales M., 2004, Ph.D Thesis, University of North Carolina
\bibitem[\protect\citeauthoryear{Lopez-Morales et al.}{2006}]{lm2006}
  Lopez-Morales M., Orosz J.A., Shaw J.S., Havelka L.,
  Arevalo M.J., McIntyre T., Lazaro, C. , 2006, ApJ, submitted ({\tt astro-ph/0610225})
\bibitem[\protect\citeauthoryear{Lucy}{1967}]{lucy67}
  Lucy L.B., 1967, Zeitschrift f\"{u}r Astrophysik, 65, 89
\bibitem[\protect\citeauthoryear{Mazeh et al.}{2002}]{maz2002}
  Mazeh T., Prato L., Simon M., Goldberg E., Norman D., Zucker S.,
  2002, ApJ, 564, 1007
\bibitem[\protect\citeauthoryear{Mullan \& MacDonald}{2001}]{mm2001}
  Mullan D.J., MacDonald J., 2001, ApJ, 559, 353
\bibitem[\protect\citeauthoryear{Nelson \& Davis}{1972}]{nd72}
  Nelson B., Davis W.D., 1972, ApJ, 174, 617
\bibitem[\protect\citeauthoryear{Palla \& Stahler}{1999}]{ps99}
  Palla F., Stahler S.W., 1999, ApJ, 525, 772
\bibitem[\protect\citeauthoryear{Palla et al.}{2007}]{palla2007}
  Palla F., Randich S., Pavlenko V., Flaccomio E., Pallavicini R.,
  2007, ApJ, 659, 41
\bibitem[\protect\citeauthoryear{Pont et al.}{2005}]{pont2005}
  Pont F., Bouchy F., Melo C., Santos N.C., Mayor M., Queloz D., Udry
  S., 2005, A\&A, 438, 1123
\bibitem[\protect\citeauthoryear{Popper \& Etzel}{1981}]{pe81}
  Popper D.M., Etzel P.B., 1981, AJ, 86, 102
\bibitem[\protect\citeauthoryear{Press et al.}{1992}]{press93}
  Press W. H., Teukolsky S. A., Vetterling, W. T., Flannery B. P.,
  1992, Numerical Recipes in Fortran 77: The Art of Scientific
  Computing. Cambridge Univ. Press, Cambridge, p. 402
\bibitem[\protect\citeauthoryear{Rebull et al.}{2006}]{rebull06}
  Rebull L.M., Stauffer J.R., Megeath S.T., Hora J.L., Hartmann L.,
  2006, ApJ, 646, 297
\bibitem[\protect\citeauthoryear{S\'{e}gransan et al.}{2003}]{seg2003}
  S\'{e}gransan D., Kervella P., Forveille T., Queloz D., 2003, A\&A,
  397, 5
\bibitem[\protect\citeauthoryear{Sicilia-Aguilar et al.}{2005}]{sa2005}
  Sicilia-Aguilar A., et al., 2005, AJ, 129, 363
\bibitem[\protect\citeauthoryear{Siess, Forestini \& Dougados}{Siess
    et al.}{1997}]{siess97}
  Siess L., Forestini M., Dougados C., 1997, A\&A, 324, 556
\bibitem[\protect\citeauthoryear{Siess, Dufour \& Forestini}{Siess et
    al.}{2000}]{siess2000}
  Siess L., Dufour E., Forestini M., 2000, A\&A, 358, 593
\bibitem[\protect\citeauthoryear{Simon \& Sturm}{1994}]{ss94}
  Simon K.P., Sturm E., 1994, A\&A, 281, 286
\bibitem[\protect\citeauthoryear{Southworth, Maxted \&
  Smalley}{Southworth et al.}{2004a}]{south2004a}
  Southworth J., Maxted P.F.L., Smalley B., 2004, MNRAS, 351, 1277
\bibitem[\protect\citeauthoryear{Southworth et al.}{2004b}]{south2004b}
  Southworth J., Zucker S., Maxted P.F.L., Smalley B., 2004, MNRAS, 355, 986
\bibitem[\protect\citeauthoryear{Southworth et al.}{2005}]{south2005}
  Southworth J., Smalley B., Maxted P.F.L., Claret A., Etzel P.B.,
  2005, MNRAS, 363, 529
\bibitem[\protect\citeauthoryear{Stassun et al.}{1999}]{smmv99}
  Stassun K.G., Mathieu R.D., Mazeh T., Vrba F.J., 1999, AJ, 117, 2941
\bibitem[\protect\citeauthoryear{Stassun et al.}{2002}]{st2002}
  Stassun K.G., van den Berg M., Mathieu R.D., Verbunt F., 2002, A\&A,
382, 899S
\bibitem[\protect\citeauthoryear{Stassun et al.}{2004}]{st2004}
  Stassun K.G., Mathieu R.D., Vaz L.P.R., Stroud N., Vrba F.J., 2004,
  ApJS, 151, 357
\bibitem[\protect\citeauthoryear{Stassun, Mathieu \& Valenti}{Stassun et al.}{2006}]{st2006}
  Stassun K.G., Mathieu R.D., Valenti J.A., 2006, Nature, 440, 311
\bibitem[\protect\citeauthoryear{Stassun, Mathieu \& Valenti}{Stassun et al.}{2007}]{st2007}
  Stassun K.G., Mathieu R.D., Valenti J.A., 2007, ApJ, accepted ({\tt astro-ph/0704.3106})
\bibitem[\protect\citeauthoryear{Tonry \& Davis}{1979}]{td79}
  Tonry J., Davis M., 1979, AJ, 84, 1511
\bibitem[\protect\citeauthoryear{Tody}{1993}]{tody93}
  Tody D., 1993, in ASP Conf. Series, Vol 52, p. 173, Astronomical
  Data Analysis Software and Systems II, eds. R.J. Hanisch,
  R.J.V. Brissenden, J. Barnes
\bibitem[\protect\citeauthoryear{Yi et al.}{2001}]{yi2001}
  Yi S., Demarque P., Kim Y.-C., Lee Y.-W., Ree C.H., Lejeune T.,
  Barnes S., 2001, ApJS, 136, 417
\end{thebibliography}
\end{document}